\documentclass[aps,prb,twocolumn,showpacs,floatfix]{revtex4}

\usepackage{graphicx,amssymb,amsfonts,amsmath}
\usepackage [dvipdfm]{hyperref}
\usepackage{subfigure}

\begin{document}

\title{Dislocation-induced superfluidity in a model supersolid}

\author{D. Goswami, K. Dasbiswas, C.-D. Yoo and Alan T. Dorsey}

\affiliation{Department of Physics, University of Florida, P.O. Box 118440, Gainesville, FL 32611-8440}

\date{\today}

\begin{abstract}
Motivated by recent experiments on the supersolid behavior of $^4$He, we study the effect of an edge dislocation in promoting superfluidity in a Bose crystal. Using Landau theory,  we couple the elastic strain field of the dislocation to the superfluid density, and use a linear analysis to show that superfluidity nucleates on the dislocation before occurring in the bulk of the solid. Moving beyond the linear analysis, we develop a systematic perturbation theory in the weakly nonlinear regime, and use this method to integrate out transverse degrees of freedom and derive a one-dimensional Landau equation for the superfluid order parameter. We then extend our analysis to a network of dislocation lines, and derive an $XY$ model for the dislocation network by integrating over fluctuations in the order parameter. Our results show that the ordering temperature for the network has a sensitive dependence on the dislocation density, consistent with numerous experiments that find a clear connection between the sample quality and the supersolid response.
\end{abstract}
\maketitle

\section{INTRODUCTION}

In 2004 Kim and Chan \cite{kim04.1,kim04.2} discovered a remarkable anomaly in the response of $^4$He crystals to ac rotation. Their results, together with subsequent experiments, \cite{rittner06,aoki07,rittner07,kondo07,clark07,day06,day07,aoki08,penzev08,hunt09} have a compelling interpretation as the nonclassical rotational inertia (NCRI) expected for the elusive supersolid phase of matter.\cite{andreev69,leggett70,chester70} However, there are puzzles--for instance, the sensitive dependence of the Kim-Chan effect on sample preparation and quality.\cite{rittner06} Indeed, theoretical studies suggest that a pristine $^4$He crystal is ``insulating,'' in that it does not support off-diagonal long range order (Bose condensation).\cite{prokofev07} The emerging consensus is that sample defects, likely in the form of dislocations, play an important role in explaining disparate experimental results.\cite{balibar08} Most existing theoretical works have focused on single dislocations; for instance, quantum Monte Carlo studies \cite{boninsegni07} have shown the existence of superfluidity along the core of screw dislocations in a model $^4$He crystal. In a phenomenological approach, Landau models show that superfluidity nucleates first on edge dislocations in a superconductor or a Bose solid, \cite{nabutovskii78,gurevich97,toner08,dasbiswas10} and Shevchenko\cite{shevchenko87, shevchenko88}, Toner\cite{toner08}, Bouchaud and Biroli\cite{bouchaud08} have worked out some of the properties of a network of dislocations in a quantum crystal. In this paper we continue and extend the work of Shevchenko and of Toner by systematically deriving an effective random-bond $XY$ model for a network of superfluid dislocations.

This paper is organized as follows. We first introduce our model for superfluidity in the presence of a single edge dislocation, by coupling the elastic strain field due to the dislocation to the superfluid order parameter.\cite{dorsey06} Then, using a linear stability analysis, we show that superfluidity always nucleates first on the dislocation. A systematic, weakly nonlinear analysis is developed and used to derive the one-dimensional Landau theory for superfluidity along the dislocation axis. Finally, we incorporate thermal fluctuations into our description of the superfluidity, and determine the effective coupling between two ``sites'' along a single dislocation. The last result motivates an effective description of the network superfluidity in terms of a random-bond $XY$ model, and we argue that such a model has an ordering temperature exponentially sensitive to the dislocation density. A series of appendices further develops the weakly nonlinear analysis: Appendix A shows how to improve the naive weakly nonlinear analysis using the method of strained coordinates; Appendix B demonstrates the efficacy of the analysis with simple variant of the edge dislocation model; and Appendix C extends the equilibrium results to the time-dependent Ginzburg-Landau theory.

\section{Superfluid-dislocation coupling in Landau theory}

Following Dorsey, Goldbart, and Toner (DGT),\cite{dorsey06} we analyze the ordered phase using a Landau theory in which the superfluid order parameter $\psi(\mathbf{x})$ couples to the strain tensor $u_{ij}$ induced by a quenched dislocation. To simplify the analysis we assume the solid is isotropic, so the superfluid couples to the trace of the strain tensor (i.e., divergence of the displacement field), which is the fractional local volume change of the solid. For a single edge dislocation along the $z$-axis, with Burger's vector $\mathbf{b}$ along the $y$ axis, the trace of the strain tensor is \cite{landau_lifshitz_elasticity}
\begin{equation}
 u_{ii} = \frac{4 \mu}{2\mu + \lambda} \frac{b \cos\theta}{r},
\end{equation}
where $\mu$ and $\lambda$ are the Lam\'e elastic constants. We have introduced the coordinates $\mathbf{x}=(\mathbf{r},z)$, with $\mathbf{r}$ in the $x-y$ plane [$(r,\theta)$ are polar coordinates in the $x-y$ plane]. The Landau free energy functional for the isotropic supersolid is then \cite{dorsey06,toner08}
\begin{equation}
F = \int d^3 x \bigg[
 \frac{c}{2} |\nabla \psi|^2 + \frac{1}{2}a(\textbf{r}) |\psi|^2 + \frac{1}{4} u|\psi|^4
\bigg],
\label{functional1}
\end{equation}
with
\begin{equation}
a(\mathbf{r}) = a_0 \left( t_0 + B \frac{\cos\theta}{r}\right).
\end{equation}
Here $a_0$, $c$, and $u$ are phenomenological parameters (all positive); $t_0 = (T-T_0)/T_0$ is the reduced temperature, with $T_0$ the mean-field critical temperature for the supersolid transition in the \emph{absence} of the dislocation ($t_0>0$ is the normal solid, and $t_0<0$ the supersolid); and $B$ is a coupling constant (into which we have absorbed the elastic constants). Cast in this form, the coupling between the dislocation and the superfluid order parameter can be thought of as a local change in the critical temperature due to local changes in the specific volume of the solid.

To simplify the subsequent analysis, we introduce the characteristic length scale $l=c/a_0 B$, order parameter scale $\chi = a_0 B/\sqrt{c u}$, and free energy scale $F_0 = a_0 B c/2u$, and the dimensionless (primed) quantities $\mathbf{x}'= \mathbf{x}/l$, $\psi'=\psi/\chi$, and ${\cal F} = F/F_0$; in terms of the dimensionless quantities the free energy becomes
\begin{equation}
{\cal F} = \int d^3x \bigg\{
  |\nabla \psi|^2 + \left[ V(\mathbf{r}) - E\right] |\psi|^2 + \frac{1}{2} |\psi|^4
\bigg\},
\label{functional2}
\end{equation}
where $V(\mathbf{r}) = \cos\theta/r$, $E \equiv - c t_0/a_0 B^2$, and we have dropped the primes on all quantities for clarity of presentation.

\section{Linear stability analysis}\label{sec:linear}

Before proceeding further, we analyze the behavior in the high-temperature, normal phase. In this phase we can neglect the quartic term in the free energy; the resulting quadratic free energy is
\begin{eqnarray}
{\cal F}_0 & = & \int d^3x \bigg\{|\nabla \psi|^2 + \left[ V(\mathbf{r}) - E\right] |\psi|^2 \bigg\}\nonumber\\
 & = & \int d^3x\, \psi^* (\hat{H} - E)\psi ,
\label{quadratic1}
\end{eqnarray}
where the Hermitian linear operator $\hat{H}$ is given by
\begin{equation}
\hat{H} = -\nabla^2 + V(\mathbf{r}).
\end{equation}
We can diagonalize the free energy by introducing a complete set of orthonormal eigenfunctions $\Psi_n(\mathbf{x})$ of $\hat{H}$,
\begin{equation}
\hat{H} \Psi_n = E_n \Psi_n,
\label{linear_operator}
\end{equation}
where $n$ labels the states, and we assume that the eigenvalues $E_n$ are ordered such that $E_0<E_1<\ldots$. Equation (\ref{linear_operator}) is equivalent to a Schr\"odinger equation for a particle in a potential $V(\mathbf{r})$, and for the dipolar potential $V(\mathbf{r}) = \cos\theta/r$ the spectrum and eigenfunctions were obtained with extensive numerical work in Ref.~\onlinecite{dasbiswas10}. Expanding the order parameter in terms of the eigenfunctions,
\begin{equation}
\psi(\mathbf{x}) = \sum_n A_n \Psi_n(\mathbf{x}),
\end{equation}
with expansion coefficients $A_n$, and substituting into the free energy, after using the orthogonality properties of the eigenfunctions we obtain
\begin{equation}
{\cal F}_0=\sum_n (E_n - E) |A_n|^2.
\label{quadratic2}
\end{equation}
The free energy is positive as long as $E_n > E$, for all $n$. Recall that $E \equiv - (c/a_0 B^2) (T-T_0)/T_0$ (with $a_0$ and $c$ both positive), so high temperatures $T$ correspond to large, negative values of $E$. As we decrease $T$, $E$ increases, until eventually we hit a \emph{condensation} temperature $T_\mathrm{cond}$ at which $E(T_\mathrm{cond}) = E_0$; below this temperature the quadratic free energy ${\cal F}_0$ becomes unstable (negative). Rearranging a bit, we have
\begin{equation}
\frac{T_\mathrm{cond} - T_0}{T_0} = - E_0 \frac{a_0 B^2}{c}.
\end{equation}
If $\hat{H}$ has negative eigenvalues--i.e., if the equivalent Schr\"odinger equation has bound states--then $T_\mathrm{cond}>T_0$, and the dislocation induces superfluidity \emph{above} the bulk ordering temperature. As emphasized in Refs.~\onlinecite{nabutovskii78,toner08,dasbiswas10}, the dipolar potential $\cos\theta/r$, which has an attractive region irrespective of the coupling constants, \emph{always} has a negative energy bound state. We are thus lead to the surprising and important conclusion \cite{nabutovskii78,toner08} that superfluidity first nucleates around the edge dislocation before appearing in the bulk of the material.

Just below the condensation temperature $T_\mathrm{cond}$, the nucleated order parameter has the form
\begin{equation}\label{ground_state}
\psi = A_0 \Psi_0(\mathbf{r}),
\end{equation}
where $\Psi_0$ is the normalized ground state wavefunction and $A_0$ is an amplitude that is fixed by the nonlinear terms in the free energy.\cite{z_dependence} Substituting Eq.~(\ref{ground_state}) into the dimensionless free energy, Eq.~(\ref{functional2}), we obtain
\begin{equation}\label{}
{\cal F} =(E_0 - E) |A_0|^2 + \frac{1}{2} g |A_0|^4,
\end{equation}
where the coupling constant $g$ is given by
\begin{equation}\label{g}
g = \int d^2r\, \Psi_0^4(\mathrm{r}).
\end{equation}
Minimizing the free energy with respect to $A_0$, we obtain
\begin{equation}\label{}
A_0 = \sqrt{(E - E_0)/g},
\end{equation}
and the minimum value of the free energy is
\begin{equation}\label{}
{\cal F}_\mathrm{min} = - \frac{(E-E_0)^2}{2g}.
\end{equation}
From the extensive numerical work of Dasbiswas \emph{et al}.\cite{dasbiswas10} we know that for the dipole potential the ground state energy is $E_0=-0.139$ (with the energy of the first excited state $E_1 = -0.0414$), and the coupling constant $g=0.0194$.

\begin{figure}[ht]
 \centering
 \includegraphics[width = 3in, height = 4in]{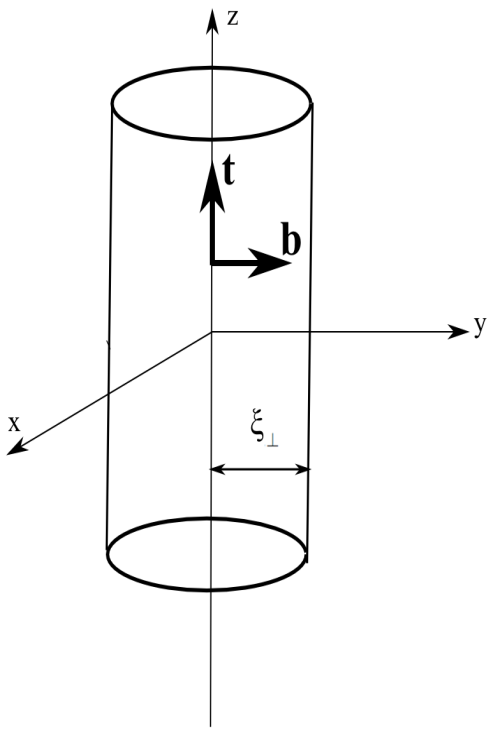}
  \caption{Schematic showing the dislocation axis (along $z$)  and the tubular superfluid region that develops around it. The radius of the cylinder is determined by the length scale of the ground state wavefunction. The axis of the tube will be offset from the dislocation axis.}
  \label{tube}
\end{figure}

To recap--following the work of previous authors,\cite{nabutovskii78,toner08} we have shown that superfluidity always nucleates \emph{first} on edge dislocations, and we have calculated the form of the order parameter near $E_0$, the threshold value of $E$ (i.e., at temperatures just below the condensation temperature). Physically, we imagine a cylindrical tube of superfluid, with a radius equal to the transverse correlation length (of order 1 in our dimensionless units), that encircles the dislocation (shown schematically in Fig.~\ref{tube}). However, this naive mean-field picture ignores the thermal fluctuations which destroy the one-dimensional superfluidity on long length scales. What is needed is an effective one-dimensional model for the superfluid, capable of capturing nontrivial fluctuation effects. In the next section we derive this one-dimensional model using a systematic, weakly-nonlinear analysis near the threshold $E_0$.

\section{Weakly nonlinear analysis near threshold}\label{sec:nonlinear}

Within the mean-field Landau theory, the order parameter configurations that minimize the free energy are solutions to the Euler-Lagrange equation
\begin{equation}\label{nonlinear}
\frac{\delta {\cal F}}{\delta \psi^*} = 0 = -\nabla^2\psi + \left[ V(\mathbf{r}) - E\right] \psi + |\psi|^2\psi.
\end{equation}
This nonlinear field equation is difficult to solve, even numerically. Instead, we resort to a \emph{weakly nonlinear analysis} \cite{drazin_reid} near the threshold for the linear instability. Our goal is to ``integrate out" the modes transverse to the dislocation and obtain an effective model for the one-dimensional superfluid nucleated along the dislocation. We then treat thermal fluctuations of this one-dimensional superfluid in the next Section.

We start by introducing a control parameter
\begin{equation}\label{}
\epsilon \equiv E - E_0
\end{equation}
that measures distance from the linear instability. From the analysis in the preceding Section, we see that the order parameter near threshold scales as $\epsilon^{1/2}$, which suggests a rescaling of the order parameter
\begin{equation}\label{}
\psi = \epsilon^{1/2} \phi,
\end{equation}
with $\phi$ a quantity whose amplitude is ${\cal O}(1)$.
Next, note that if we had included the plane wave behavior along the $z$-axis in the analysis in Sec.~\ref{sec:linear}, the coefficient of $|A_n|^2$ in the quadratic free energy, Eq.~(\ref{quadratic2}), would be $-\epsilon + k^2$. This suggests the important fluctuations along the $z$-axis occur at wavenumbers $k \sim \epsilon^{1/2}$, or at length scales of order $\epsilon^{-1/2}$ (i.e., long wavelength fluctuations are important close to threshold); the required rescaling is
\begin{equation}\label{}
z=\epsilon^{-1/2} \zeta.
\end{equation}
Substituting these variable changes into Eq.~(\ref{nonlinear}), and writing $E = E_0 + \epsilon$, we obtain
\begin{equation}\label{}
\hat{L} \phi = \epsilon \left[\partial^2_\zeta \phi + \phi - |\phi|^2\phi\right],
\end{equation}
where the Hermitian linear operator $\hat{L}$ is given by
\begin{equation}\label{L}
\hat{L} = -\nabla^2_\perp + V(\mathbf{r}) - E_0,
\end{equation}
with $\nabla^2_\perp$ the Laplacian in dimensions transverse to $z$.
Next, we expand $\phi$ in powers of $\epsilon$,
\begin{equation}\label{}
\phi = \phi_0 + \epsilon \phi_1 + \epsilon^2 \phi_2 + \ldots .
\end{equation}
Collecting terms, we obtain the following hierarchy of equations:
\begin{eqnarray}\label{zero_order}
{\cal O}(1):& & \qquad \hat{L} \phi_0 = 0, \nonumber\\
{\cal O}(\epsilon): & & \qquad \hat{L} \phi_1 = \partial^2_\zeta \phi_0 + \phi_0 - |\phi_0|^2\phi_0, \\
{\cal O}(\epsilon^2): & & \qquad \hat{L} \phi_2 = (1-3\phi_0^2)\phi_1. \nonumber
\end{eqnarray}
The solution of the ${\cal O}(1)$ equation is the normalized ground state eigenfunction, $\Psi_0(\mathbf{r})$; there is an overall integration constant $A_0(\zeta)$, so that
\begin{equation}\label{}
\phi_0 = A_0(\zeta) \Psi_0(\mathbf{r}).
\end{equation}
Substitute this into the right hand side of the ${\cal O}(\epsilon)$ equation,
\begin{equation}\label{}
\hat{L} \phi_1 = \Psi_0\partial^2_\zeta A_0 + \Psi_0 A_0 - \Psi_0^3 |A_0|^2A_0  .
\end{equation}
We can determine $A_0$ by left multiplying this equation by $\Psi_0$, integrating on $d^2 r$, and using the fact that $\hat{L}$ is Hermitian, to find
\begin{equation}\label{A0}
\partial^2_\zeta A_0 + A_0 - g |A_0|^2 A_0 = 0,
\end{equation}
where $g$ is defined in Eq.~(\ref{g}). This is the \emph{solvability condition} for the existence of nontrivial solutions of the ${\cal O}(\epsilon)$ equation.
In principle we could solve this equation for $A_0$, substitute back into the right hand side of the ${\cal O}(\epsilon)$, and solve the resulting inhomogeneous equation to obtain $\phi_2$. In practice this is difficult for the dipole potential, so we will stop at this level of the perturbation theory.

We can recast Eq.~(\ref{A0}) in terms of $z=\epsilon^{-1/2}$ and $\varphi = \epsilon^{1/2} A_0$ as
\begin{equation}\label{}
\partial^2_z \varphi(z) + \epsilon \varphi - g |\varphi|^2\varphi = 0,
\end{equation}
which is the Euler-Lagrange equation for the free energy functional
\begin{equation}\label{}
{\cal F} = \int dz\, \left[ \frac{1}{2} |\partial_z \varphi|^2 - \frac{1}{2} \epsilon |\varphi|^2 + \frac{g}{4}|\varphi|^4\right].
\end{equation}
Reinstating the dimensions, we obtain
\begin{equation}\label{}
F = \int dz\, \left[ \frac{c}{2} |\partial_z \varphi|^2 + \frac{a}{2}|\varphi|^2 + \frac{b}{4}|\varphi|^4\right],
\end{equation}
where $a = a_0 t$, $t= (T - T_\mathrm{cond})/T_\mathrm{cond}$ is the reduced temperature measured relative to the \emph{condensation} temperature, and $ b = g u$. To summarize, we have integrated out the transverse degrees of freedom (the fluctuations of which have a nonzero energy gap) and reduced the full three-dimensional problem to an effective one-dimensional model. In the next section we will study the fluctuations of this one-dimensional model, and derive an effective phase-only model for a dislocation network.

A deficiency of our perturbative approach is the appearance of the term $\phi_0$ on the right hand side the ${\cal O(\epsilon)}$ equation in Eq.~(\ref{zero_order}). Since $\phi_0$ is a zero mode of $\hat{L}$, it becomes a secular term in subsequent orders of the perturbative expansion. For some systems--e.g., a nonlinear oscillator--such secular terms lead to unbounded solutions that require regularization using a more sophisticated multiple-scale analysis.\cite{kevorkian96} In our application the secular terms are benign, as the solution decays at infinity (but with the wrong decay rate). However, in Appendix A we show that the secular terms can be eliminated using the \emph{method of strained coordinates},\cite{kevorkian96} with results that, on the whole, are equivalent to the results above (but with an improved approximation to the order parameter). In Appendix B we apply our perturbative methods to the two-dimensional Coulomb potential $V(\mathbf{r}) = -1/r$, for which the calculations can be explicitly worked out through ${\cal O}(\epsilon^2)$, and we show that the results are in close agreement with detailed numerical solutions of the nonlinear field equation. Finally, in Appendix C we show that the same methods may be used to treat dynamic equations of motion for the order parameter.

\section{Derivation of the dislocation network model}

So far we have considered a single edge dislocation, reducing the full three-dimensional Landau theory to an effective one-dimensional theory for a superfluid tube localized near the dislocation core. However, a $^4$He crystal will consist of a tangle of dislocations, many of which will cross when they come within a transverse correlation length of each other. Conceptually, we can model this as a random lattice (or network) of dislocations, with the crossing points serving as lattice sites. While thermal fluctuations destroy any long range order in a single, one-dimensional tube, the lattice of tubes will generally order at a temperature characteristic of the phase stiffness between adjacent lattice sites. This is the motivation behind the models developed by Shevchenko \cite{shevchenko87} and Toner \cite{toner08}. In this Section we revisit the Shevchenko and Toner models with a systematic approach to calculating the coupling between adjacent lattice sites in the network model, and obtain new results on the length scaling of the coupling constant. We conclude with some observations regarding implications of our results for experiments on the putative supersolid phase of $^4$He.

We start by considering the correlations between two points along a single superfluid tube, separated by a distance $L$. Using the notation of the previous section, $\varphi(0)=\varphi_1$ and $\varphi(L)=\varphi_2$ are the values of the superfluid order parameter at two sites along the tube. The correlations are captured by the propagator $K(\varphi_2, \varphi_1;L)$, which can be obtained from the functional integral
\begin{widetext}
\begin{equation}
K(\varphi_2, \varphi_1;L) = \exp(-\beta H_\mathrm{eff}) = \int_{\varphi(0)=\varphi_1}^{\varphi(L)=\varphi_2} {\cal D}(\varphi, \varphi^*) \exp \Bigg\{-\beta\int^L_0 dz\bigg[ \frac{c}{2}|\partial_z \varphi |^2 + \frac{a}{2}|\varphi|^2 + \frac{b}{4} |\varphi|^4 \bigg] \Bigg\},
\label{pathintegral}
\end{equation}
\end{widetext}
where $a$, $b$ and $c$ are the parameters of the one-dimensional Landau theory derived in the previous section, and $H_\mathrm{eff}$ is the effective Hamiltonian that characterizes the coupling between the lattice sites at $z=0$ and $z=L$. Cast in this form, we see that the functional integral for the classical one-dimensional system is equivalent to the Feynman path integral for a quantum particle in a two-dimensional quartic potential (two-dimensional because the order parameter $\varphi$ is complex), with $z$ in the classical problem replaced by the imaginary time $\tau$ for the quantum system.\cite{feynman_hibbs} Indeed, previous authors have used this analogy to study the effect of thermal fluctuations on the resistive transition in one-dimensional superconductors,\cite{gruenberg72, scalapino72} obtaining essentially exact results for the correlation length and thermodynamic properties. Consistent with the Mermin-Wagner theorem,\cite{mermin66} these authors find no singularities in the thermodynamic properties, and a one-dimensional correlation length that grows as the temperature is reduced, but never diverges.\cite{scalapino72} These important features are conveniently captured using a simple Hartree approximation,\cite{tucker71} in which the quartic term is absorbed into the quadratic term with the quadratic coefficient redefined as
\begin{equation}
\bar{a} = a + \frac{1}{2} b\langle|\varphi|^2\rangle ,
\label{hf}
\end{equation}
where $a= a_0 (T-T_\mathrm{cond})/T_\mathrm{cond}$ and $\langle|\varphi|^2\rangle$ a statistical average with respect to the effective quadratic theory. Carrying out the averaging, we obtain \cite{tucker71}
\begin{equation}
\langle|\varphi|^2\rangle = \frac{k_B T}{c \xi_\perp^2} \xi,
\label{mfsol}
\end{equation}
where $\xi = (c/\bar{a})^{1/2}$ is the correlation length for the one-dimensional system along the superfluid tube, and $\xi_{\perp}$ is the cross-sectional dimension of the 1D superfluid region (the transverse correlation length, as shown in Fig.~\ref{tube}). Inserting this result into the Hartree expression of Eq.~(\ref{hf}), we obtain a cubic equation for the correlation length,
\begin{equation}
\frac{1}{\xi^2} = \frac{1}{\xi_0^2} + \left(\frac{k_B T b}{2c^2\xi_\perp^2}\right) \xi,
\label{cubeqn}
\end{equation}
where $\xi_0 = (c/a)^{1/2}$ is the Gaussian correlation length. While $\xi_0$ diverges at $T_\mathrm{cond}$, $\xi$ remains finite at all temperatures (growing as the temperature is lowered),\cite{scalapino72} reflecting the lack of long-range order in the one-dimensional superfluid tube.

Continuing with the Hartree approximation, we can find the explicit form of the propagator by exploiting an analogy with the partition function for a two-dimensional quantum harmonic oscillator; the result is\cite{feynman_hibbs}
\begin{widetext}
\begin{equation}
K(\varphi_2, \varphi_1;L) = k(L) \exp \Bigg\{ -\frac{\beta c}{2 \xi \sinh( L/\xi )} \bigg[(|\varphi_2|^2 + |\varphi_1|^2)\cosh(L/\xi) - 2|\varphi_2||\varphi_1|\cos(\theta_1 - \theta_2)\bigg] \Bigg\},
\label{propagator}
\end{equation}

\end{widetext}
where the prefactor $ k(L) = \beta c / 2 \pi \xi \sinh(L/\xi)$, and $\varphi_1 = |\varphi_1|e^{i\theta_1}$ and $\varphi_2 = |\varphi_2|e^{i\theta_2}$.
The effective Hamiltonian (up to an additive constant) is given by
\begin{equation}
H_\mathrm{eff} = \frac{c}{2\xi}\coth(L/\xi)(|\varphi_2|^2 + |\varphi_1|^2) - J_{12}(L)\cos(\theta_1 - \theta_2),
\label{Hamiltonian}
\end{equation}
where
\begin{eqnarray}
J_{12}(L) &= &\frac{c |\varphi_2||\varphi_1|}{\xi \sinh(L / \xi)} \nonumber\\
 &=&  \left\{
     \begin{array}{ll}
      c|\varphi_2||\varphi_1|/L, \quad L/\xi\ll 1; \\
      (2c |\varphi_2||\varphi_1|/\xi) e^{-L/\xi}, \quad  L/\xi\gg 1.
      \end{array}
      \right.
\label{J}
\end{eqnarray}
The last term in Eq.~(\ref{Hamiltonian}) is the one of interest, as it couples the phases at the neighboring sites through an effective ``ferromagnetic'' coupling constant $J_{12}>0$. The behavior of $J_{12}$ as a function of the inter-site separation $L$ is one of our important results--for small $L/\xi$, $J_{12}\sim 1/L$, reproducing the result of Toner,\cite{toner08} whereas for large $L/\xi$, $J_{12}\sim e^{-L/\xi}$. Since $\xi$ is finite for all $T$, for a sufficiently dilute network of dislocations we will always satisfy the latter condition--i.e., dilute networks of dislocations possess coupling constants exponentially small in the dislocation density. 

We emphasize that a one-dimensional system with a continuous symmetry does not exhibit long range order--order parameter correlations decay exponentially on the scale of the correlation length $\xi$. As a result, we cannot replace the full Landau functional with a phase-only model (in which one assumes a well-formed order parameter amplitude), and we need to treat the phase and amplitude fluctuations on the same footing, in the spirit of the work by Scalapino {\it{et al.}}\cite{scalapino72}. In this respect our results differ from Toner's \cite{toner08}, who used a phase-only treatment and found a coupling constant that scales as $1/L$. Toner's result \emph{does} apply at short length scales ($L\ll \xi$), where there is local superfluid order and a phase-only approximation can be used. On the other hand, at long length scales ($L\gg \xi$) the exponential decay of correlations results in a coupling constant that is exponentially small in $L$. The result in Eq. (\ref{J})correctly captures both the small and large distance behavior of the coupling constant. 

So far we have systematically derived an effective Hamiltonian that describes the phase coupling between two points (lattice sites) on a single dislocation. To make the conceptual leap to the dislocation network, we propose that the appropriate model for the network is a random bond $XY$ model of the form
\begin{equation}\label{}
H = - \sum_{\langle ij\rangle} J_{ij} \cos (\theta_i - \theta_j),
\end{equation}
where $\langle ij\rangle$ represents nearest-neighbor lattice sites, and $J_{ij}$ is a (positive) coupling between the sites that scales as $e^{-L_{ij}/\xi}$ for a sufficiently dilute network of dislocations. As noted by Toner,\cite{toner08} the randomness is irrelevant in the renormalization-group sense,\cite{harris74} and in three dimensions we expect the superfluidity in the network to order when the temperature is of order the typical coupling strength $[J_{ij}]$; i.e., $k_B T_c ={\cal O}([J_{ij}] )$. Again, for a dilute network of dislocations (with areal dislocation density $n\sim 1/L^2$), we would find $T_c\sim e^{-1/(n^{1/2}\xi)}$, a remarkably sensitive dependence on the dislocation density.

How do these theoretical findings compare with the experiments? The emerging consensus is that the putative supersolid response depends on sample quality and preparation (for instance, see Rittner and Reppy\cite{rittner06}). As an example, the NCRI fraction shows a sensitive dependence on the sample quality (varying from as much as $20\%$ for samples prepared by the blocked capillary method and thus having more disorder to $0.5\%$ for those prepared under constant pressure, \cite{rittner07}), while the onset temperature itself shows a weak dependence on disorder.  This could suggest that existing experiments are in the high dislocation density regime ($L\ll \xi$) where the coupling constant (and therefore the critical temperature) scale as $1/L$. Unfortunately, there have been no systematic studies that correlate the dislocation density of solid $^4$He samples with the onset temperature for supersolid signal.  Ultrasound attenuation experiments on solid $^4$He \cite{iwasa79,iwasa80} suggested dislocation densities ranging from 10$^{4}$-10$^{6}$ cm$^{-2}$; at present the dislocation density is not even known to within an order of magnitude, which makes it difficult to predict $T_c$ from our theory. Moreover, the Landau theory parameters for solid $^4$He are unknown (unlike the situation in conventional superconductors, for example), compounding the difficulties in directly comparing our results to the existing experiments. However, the dislocation network model still serves as a useful conceptual starting point for understanding some aspects of dislocations on Bose crystals, and our work is important in establishing the equivalence between the network model and a more fundamental Landau theory. Finally, we note that our theory does not include dynamics (however, see Appendix C for a rudimentary discussion of dynamics) and we cannot comment on experiments that study the rate of superflow, such as by Ray and Hallock \cite{hallock08}. However we believe that the model proposed above serves as a useful starting point to numerically simulate and understand some of these effects.

\section{Summary}

We have constructed a model for a supersolid based on superfluidity induced along a network of dislocations. Starting from a Ginzburg-Landau theory for the bulk solid, we are able to systematically derive a one-dimensional equation describing superfluidity along a single dislocation. Then we complete the picture by considering a network of dislocations and the effect of overlap of these strands of 1D superfluid, that give us back superfluid behavior for the bulk.  In passing, we note the same effect should be observable in superconductors, where it should be possible to calculate the various constants in the Landau theory in terms of microscopic quantities (something we cannot do for solid $^4$He). One of our more striking results is the sensitive dependence of the transition temperature for the dislocation network on the dislocation density (i.e., the sample quality).

\begin{acknowledgments}
The authors are indebted to John Toner for useful discussions. A.T.D. would like to thank the Aspen Center for Physics, where part of this work was completed. This work was supported by the National Science Foundation, Grant No. DMR-0705690.
\end{acknowledgments}

\appendix

\section{Method of strained coordinates}

In this Appendix we show how to improve upon the naive perturbation theory developed in Sec. IV, using the \emph{method of strained coordinates}. We start with the Euler-Lagrange equation [see Eq.~(\ref{nonlinear})]
\begin{equation}\label{nonlinear1}
    -\nabla^2 \phi + V(\mathbf{x})\phi - E\phi + \epsilon |\phi|^2\phi = 0,
\end{equation}
where $E = E_0 + \epsilon$, with $E_0$ the lowest eigenvalue of $-\nabla^2 + V(\mathbf{x})$ (the ground state energy). Assuming $V\rightarrow 0 $ as $|\mathbf{x}|\rightarrow \infty$, the bound solutions of this equation behave as $\phi \sim \exp( - \sqrt{-E} |\mathbf{x}|)$ for $|\mathbf{x}|\rightarrow \infty$; there is a characteristic length scale $1/\sqrt{-E}$. To account for this scale, define new coordinates $X_i = x_i/l$, where $l=1/\sqrt{1+\epsilon/E_0}$. In terms of these new coordinates, we have
\begin{equation}\label{}
-(1 + \epsilon/E_0) \nabla^2_X \phi + V(l\mathbf{X})\phi - E \phi + \epsilon |\phi|^2 \phi = 0.
\end{equation}
For small $\epsilon$ the length scale is $l = 1 - \epsilon/2E_0 + (3/8)(\epsilon/E_0)^2 + \ldots$; the potential is then
\begin{equation}\label{}
V(l \mathbf{X}) \approx V(\mathbf{X}) - \frac{\epsilon}{2E_0} \mathbf{X}\cdot\nabla_X V + {\cal O}(\epsilon^2).
\end{equation}
Introducing the linear operator
\begin{equation}\label{linear1}
\hat{L} = - \nabla^2_X + V(\mathbf{X}) - E_0,
\end{equation}
and rearranging things a bit, we have
\begin{equation}\label{nonlinear2}
\hat{L}\phi = -\frac{\epsilon}{E_0} \hat{L}\phi +\frac{\epsilon}{E_0}\left( \frac{1}{2} \mathbf{X}\cdot\nabla V + V\right)\phi - \epsilon |\phi|^2 \phi + {\cal O}(\epsilon^2).
\end{equation}
Now expand $\phi$:
\begin{equation}\label{}
\phi = \phi_0 + \epsilon \phi_1 + {\cal O}(\epsilon^2).
\end{equation}
Substituting into Eq.~(\ref{nonlinear2}), we obtain:
\begin{eqnarray}\label{}
{\cal O}(1):\hat{L} \phi_0 &= &0,\\ \nonumber
{\cal O}(\epsilon): \hat{L} \phi_1 &=& \frac{1}{E_0}\left( \frac{1}{2} \mathbf{X}\cdot\nabla V + V\right)\phi_0 - |\phi_0|^2\phi_0 .
\end{eqnarray}
Notice that the right hand sides of both terms do not contain $\phi_0$ by itself--there are no secular terms, unlike the ``naive'' version of the perturbation theory.

The solution to the ${\cal O}(1)$ equation is $\phi_0 = A \Psi_0(\mathbf{X})$, where $\Psi_0$ is the normalized ground state eigenfunction and $A$ is an amplitude. Substituting this into the ${\cal O}(\epsilon)$ equation,
\begin{equation}\label{}
\hat{L} \phi_1 = A \frac{1}{E_0} \left( \frac{1}{2} \mathbf{X}\cdot\nabla V + V\right) \Psi_0 - A^3 \Psi_0^3.
\end{equation}
Now left multiply by $\Psi_0$, and use the fact that $\hat{L}$ is Hermitian:
\begin{equation}\label{solvability}
\langle \Psi_0,\hat{L}\phi_1\rangle = 0 = A \frac{1}{E_0} \left( \frac{1}{2} \langle\mathbf{X}\cdot\nabla V\rangle_0 + \langle V \rangle_0 \right) - g A^3,
\end{equation}
where
\begin{equation}\label{}
g = \int d^2X \Psi_0^4.
\end{equation}

Note that $\mathbf{X}\cdot\nabla V = - \mathbf{X}\cdot \mathbf{F}$, with $\mathbf{F}$ the force associated with the potential $V$. The quantum version of the virial theorem\cite{shankar} states that $\langle \mathbf{X}\cdot\mathbf{F}\rangle_0 = -2 \langle T \rangle_0$, with $T$ the kinetic energy; therefore, the quantity inside the parentheses in Eq.~(\ref{solvability}) is the ground state energy $E_0 = \langle T \rangle_0 + \langle V \rangle _0$, and we have simply
\begin{equation}\label{}
 0 = A - g A^3,
\end{equation}
or $A = 1/\sqrt{g}$.

Pulling all of the results together, we have
\begin{equation}\label{}
\phi(\mathbf{x}) = g^{-1/2} \Psi_0(\mathbf{x}/l),
\end{equation}
where $l = 1/\sqrt{1 + \epsilon/E_0}= \sqrt{E_0/E}$. The strained coordinate calculation introduces an $\epsilon$ dependence into the order parameter itself (the ``naive'' perturbation theory gave the same result, but with $l = 1$), which guarantees the correct asymptotic behavior of the order parameter.

\section{Analysis of a Landau model with a $1/r$ potential}

In this Appendix we solve a simplified version of the dipole potential, replacing $\cos\theta/r$ by the attractive two-dimensional Coulomb potential $V({\bf r}) = -1/r$. These two potentials share the same length scaling; however, the Coulomb potential is rotationally symmetric and the linear problem can be solved exactly. The details of the perturbation calculation follow the general scheme developed in Sec.~\ref{sec:nonlinear}. We compare the results of the perturbation theory with numerical solutions of the nonlinear field equation, and find close agreement for a wide range of $\epsilon$.

The energy eigenvalues for the Hamiltonian $\hat{H} = -\nabla^2_\perp - 1/r$ are given by\cite{zaslow67,yang91}
\begin{equation}\label{coulomb_spectrum}
E_n = - \frac{1}{(2n+1)^2},\quad n=0,1,2,\ldots,
\end{equation}
so $E_0 = -1$, with a ground state eigenfunction $\Psi_0(r) = \sqrt{2/\pi} e^{-r}$. The related linear operator [see Eq.~(\ref{L})] is
\begin{equation}\label{}
\hat{L} = -\nabla^2_\perp -\frac{1}{r} + 1 .
\end{equation}
Expanding $\phi$ as before,
\begin{equation}\label{expansion}
\phi = \phi_0 + \epsilon \phi_1 + \epsilon^2 \phi_2 + \ldots ,
\end{equation}
we have
\begin{equation}\label{order_zero}
\hat{L} \phi_0 = 0,
\end{equation}
with the solution
\begin{equation}\label{}
\phi_0(r) = A_0 \left(\frac{2}{\pi}\right)^{1/2} e^{-r}
\end{equation}
(note that we will ignore the $z$-dependence in this Appendix). Substituting into the left hand side of the ${\cal O}(\epsilon)$ equation,
\begin{eqnarray}\label{order_one}
  \hat{L} \phi_1  &=&  \phi_0 - |\phi_0|^2\phi_0 \nonumber \\
  &=& A_0 \left(\frac{2}{\pi}\right)^{1/2} e^{-r} - A_0^3 \left(\frac{2}{\pi}\right)^{3/2} e^{-3r}.
\end{eqnarray}
We left multiply this equation by $\sqrt{2/\pi} e^{-r}$ and integrate on $d^2 r$ to obtain $0=A_0 - (1/2\pi) A_0^3$, so that $A_0=\sqrt{2\pi}$. Substituting back into Eq.(~\ref{order_one}), and assuming $\phi_1$ has cylindrical symmetry, we obtain an inhomogeneous equation for $\phi_1$:
\begin{equation}\label{}
-\frac{1}{r} \frac{d}{dr}\left( r\frac{d\phi_1}{dr} \right) -\frac{1}{r}\phi_1 + \phi_1 = 2 e^{-r
} - 8 e^{-3r}.
\end{equation}
The explicit solution of this equation (that decays to 0 for large $r$) is
\begin{equation}\label{order1}
\phi_1(r) = c e^{-r} + e^{-3r} + r e^{-r} + \frac{1}{2}e^{-r}\left[\ln(2r) + \int_{2r}^\infty \frac{e^{-t}}{t}\, dt \right],
\end{equation}
where $c$ is an integration constant. We substitute $\phi_1(r)$ into the right hand side of the ${\cal O}(\epsilon^2)$ equation to obtain
\begin{equation}\label{order_two}
\hat{L} \phi_2 = (1-3\phi_0^2)\phi_1 .
\end{equation}
We again left multiply by $\sqrt{2/\pi} e^{-r}$, integrate on $d^2r$ and use the fact that $\hat{L}$ is Hermitian to obtain the solvability condition for the constant $c$, with the result
\begin{equation}\label{constant}
c=-\frac{11}{12} + \frac{\gamma}{2} + \ln(2) - \frac{3}{4} \ln(3)  = - 0.75887,
\end{equation}
where $\gamma=0.577210$ is the Euler-Mascheroni constant. We have now explicitly calculated \emph{\emph{two}} terms in the perturbation expansion; both terms are nonzero at the origin, with $\phi_0(0) = 2$ and $\phi_1(0) = \frac{1}{12} + \ln(2) - \frac{3}{4}\ln(3) = -0.047478$.

\begin{figure}[ht]
 \includegraphics[width = 3 in]{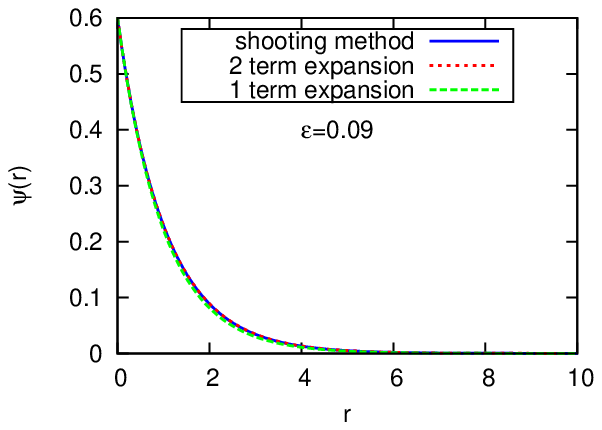}
 \includegraphics[width = 3 in]{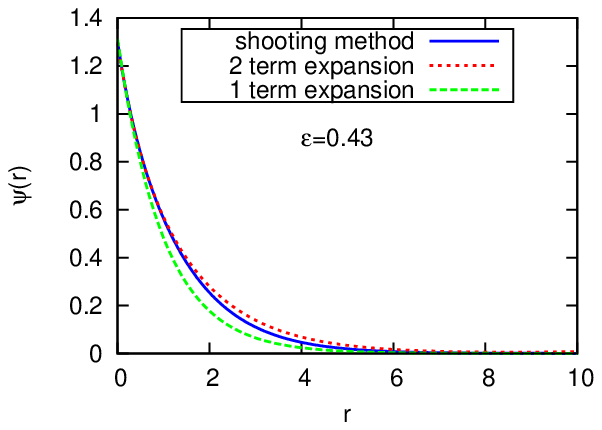}
\caption{(Color online) Order parameter $\psi(r)$ versus $r$. The blue line is the numerical solution, the green line is the one-term perturbative result, and the red line is the two-term perturbative result. The upper panel compares the results for $\epsilon=0.09$, and the lower panel for $\epsilon=0.43$. The two term expansion provides an excellent approximation to the numerical result, even for $\epsilon=0.43$.\label{fig:psi_r_versus_r}}
\end{figure}

We have solved the nonlinear field equation for the $-1/r$ potential numerically for a wide range of values of $\epsilon = E - E_0 = E+1$, using the shooting method.\cite{shooting} The results for the order parameter are presented in Fig.~\ref{fig:psi_r_versus_r} for two different values of $\epsilon$. The two-term perturbation theory gives an excellent approximation even for a fairly large value of $\epsilon=0.43$.

\begin{figure}[ht]
 \includegraphics[width = 3 in]{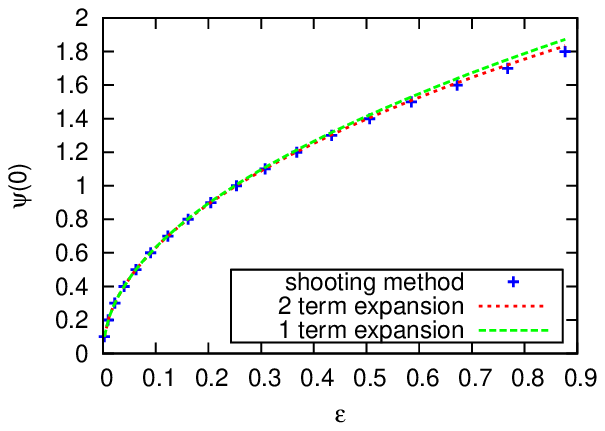}
 \includegraphics[width = 3 in]{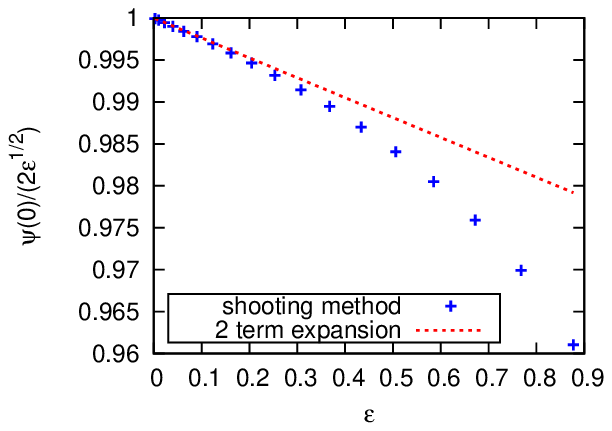}
 \caption{(Color online) Order parameter amplitude $\psi(0)$ as a function of $\epsilon$. The crosses are the numerical results, the green line is the one-term perturbative result, and the red line is the two-term perturbative result. The upper plot compares the numerical results with the perturbation theory for a wide range of $\epsilon$; the two-term perturbative result provides an excellent approximation even for values of $\epsilon$ as large as $0.8$. The lower panel is a plot of $\psi(0)/2\epsilon^{1/2}$ as a function of $\epsilon$, which highlights the role of the second order term in the expansion.\label{fig:psi(0)_versus_epsilon} }
\end{figure}

To get a sense of the efficacy of the perturbation theory, we can calculate the amplitude of the order parameter at the origin:
\begin{eqnarray}\label{}
\psi(0)& =& \epsilon^{1/2}\phi_0(0) + \epsilon^{3/2}\phi_1(0) + \ldots \nonumber \\
 & = & 2 \epsilon^{1/2} - 0.047478 \epsilon^{3/2} + \ldots\ .
\end{eqnarray}
This result is plotted in Fig.~\ref{fig:psi(0)_versus_epsilon}, along with our numerical results. Again, we see the excellent agreement between the two-term perturbation theory and the numerical results, even for relatively large values of $\epsilon$.

\begin{figure}[ht]
 \includegraphics[width = 3 in]{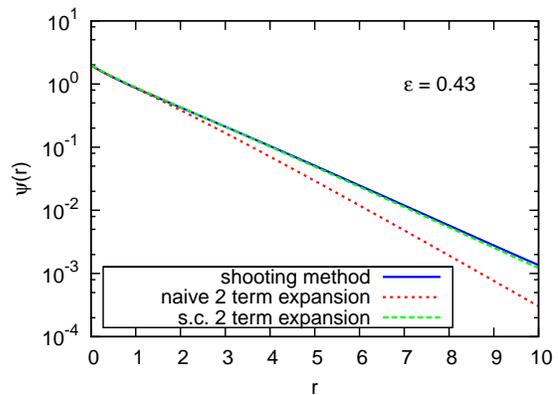}
\caption{(Color online) Comparison of the numerical shooting method (blue), naive perturbation theory (red dotted), and strained coordinates perturbation theory (green dashed) for the order parameter $\psi(r)$ for $\epsilon=0.43$. The log scale highlights the asymptotic behavior of the order parameter--the numerical and strained coordinates calculations are indistinguishable for large $r$, while the naive perturbative result clearly decays too rapidly.\label{fig:strained_coordinates}}
\end{figure}

While the naive perturbation theory does an excellent job in capturing the overall amplitude of the order parameter, it produces the wrong asymptotic behavior of the order parameter, as discussed in Appendix A. This deficiency is remedied using the method of strained coordinates, and for the sake of completeness we reanalyze the attractive Coulomb potential problem following the procedure outlined in Appendix A. As noted previously, $E_0 = -1$, so the characteristic length scale $l = (1-\epsilon)^{-1/2} = 1 - (1/2)\epsilon + (3/8)\epsilon^2 + \ldots$. Changing coordinates to $X_i=x_i/l$, and collecting terms, we have at ${\cal O}(1)$
\begin{equation}\label{}
\hat{L} \phi_0 = 0,
\end{equation}
where
\begin{equation}\label{}
\hat{L} = - \nabla^2_X - \frac{1}{R} + 1 = -\frac{1}{R}\frac{d}{dR}\left( R \frac{d}{dR}\right) - \frac{1}{R} + 1 .
\end{equation}
The solution is $\phi_0 (R) = A_0 \sqrt{2/\pi} e^{-R}$. At ${\cal O}(\epsilon)$, we have
\begin{equation}\label{}
\hat{L} \phi_1 = \frac{1}{2R} \phi_0(R) - \phi_0^3(R).
\end{equation}
Substituting $\phi_0$ into the right hand side, left multiplying by $\sqrt{2/\pi} e^{-R}$, and integrating on $d^2R$, we obtain $A_0 = \sqrt{2\pi}$, so
\begin{equation}\label{}
\phi_0(R) = 2 e^{-R}.
\end{equation}
The ${\cal O}(\epsilon)$ equation is then
\begin{equation}\label{}
-\frac{1}{R}\frac{d}{dR}\left( R \frac{d\phi_1}{dR}\right) - \frac{1}{R}\phi_1 + \phi_1   = \frac{e^{-R}}{R} - 8 e^{-3R}.
\end{equation}
The explicit solution that decays to zero as $R\rightarrow \infty$ is [one can show that integration constant $c$ is given by Eq.~(\ref{constant}) as before]
\begin{equation}\label{}
\phi_1(R) = ce^{-R} + e^{-3R} + \frac{1}{2} e^{-R} \left[ \ln(2R) + \int_{2R}^\infty \frac{e^{-t}}{t}\, dt \right] .
\end{equation}
This is almost the same result as for the naive perturbation theory obtained in Eq.~(\ref{order1}),
where $R = \sqrt{1 - \epsilon} r$. The difference is the third term $re^{-r}$ in Eq.~(\ref{order1}); this is the secular term that is generated in the naive perturbation theory. The strained coordinates method has subsumed this term into the first order term; i.e.,
\begin{equation}\label{}
2 e^{-R} = 2 e^{-\sqrt{1-\epsilon} r} \approx 2 e^{-r} + \epsilon r e^{-r} + \ldots\, .
\end{equation}
To get a sense of the improvement in capturing the correct asymptotic behavior of the order parameter, in Fig.~\ref{fig:strained_coordinates} we plot the numerical results obtained using the shooting method against the naive and strained coordinate perturbation theory. The strained coordinate result is indistinguishable from the numerical result in this log plot; the naive perturbation theory result clearly decays too rapidly for large $r$.

\section{Analysis of a time-dependent model}\label{sec:dynamics}

In this Appendix we generalize the results of Sec.~\ref{sec:nonlinear} to derive a one-dimensional dynamical model for the superfluid. For simplicity, we'll assume that there are no conserved densities, so the dynamics are described by model A (often referred to as time-dependent Ginzburg-Landau theory) in the Hohenberg-Halperin classification.\cite{hohenberg77,gp_dynamics} The relaxational equation of motion is
\begin{equation}
\frac{\partial \psi}{\partial t} = -\Gamma_0\frac{ \delta F} { \delta {\psi}^*} + \eta(\mathbf{x},t),
\end{equation}
where $\Gamma_0$ is a relaxation rate and $\eta$ is the fluctuating noise term with Gaussian white noise correlations, i.e.,
$\langle \eta({\mathbf x},t)\rangle = 0$ and $\langle \eta({\mathbf x},t)^*\eta({\mathbf x}',t')\rangle = 2k_BT\Gamma_0 \delta({\mathbf x} - { \mathbf x}') \delta(t-t')$. As before, to facilitate the reduction to a one-dimensional model we introduce dimensionless quantities, with a time scale $\tilde{t} = 2c/(\Gamma a_0^2 B^2) = 2 l^2/(\Gamma c)$. In terms of the dimensionless variables,
\begin{equation}\label{}
\frac{\partial \psi}{\partial t} = \nabla^2\psi - [V(\mathbf{r}) - E]\psi - |\psi|^2\psi + \bar{\eta}(\mathbf{x},t),
\end{equation}
where the noise correlations are given by
\begin{equation}\label{}
\langle \bar{\eta}^*(\mathbf{x},t)\bar{\eta}(\mathbf{x}',t')\rangle = 2 (k_B T/F_0) \delta({\mathbf x} - { \mathbf x'}) \delta(t-t').
\end{equation}
As before, we introduce the small parameter $\epsilon = E - E_0$, with $\psi = \epsilon^{1/2} \phi$, $z=\epsilon^{-1/2} \zeta$, $t=\epsilon^{-1}\tau$, and $\bar{\eta} = \epsilon^{3/2} \tilde{\eta}$, to obtain
\begin{equation}\label{}
\hat{L} \phi = \epsilon \left[ - \partial_\tau\phi + \partial^2_\zeta\phi + \phi - |\phi|^2\phi + \tilde{\eta}\right].
\end{equation}
We expand $\phi$ in powers of $\epsilon$, and at ${\cal O}(1)$ we have $\hat{L}\phi_0 = 0$, the solution of which is
\begin{equation}\label{}
\phi_0 = A_0(\zeta,\tau) \Psi_0(\mathbf{r}).
\end{equation}
Substituting this into the right hand side of the ${\cal O}(\epsilon)$ equation, left multiplying by $\phi_0$, and then integrating on $d^2 r$, the solvability condition yields
\begin{equation}\label{}
\partial_\tau A_0 = \partial^2_\zeta A_0 + A_0 - g |A_0|^2 A_0 + \xi,
\end{equation}
where $\xi$ is the one-dimensional fluctuating noise term (the three-dimensional term with the transverse dimensions projected out),
\begin{equation}\label{}
\xi(\zeta,\tau) = \int d^2r \Psi_0(\mathbf{r}) \tilde{\eta}(\mathbf{r},\zeta,\tau).
\end{equation}
Using the fact that $\Psi_0$ is normalized to one, it is straightforward to show that $\xi$ has Gaussian white noise correlations. Undoing the $\epsilon$ scalings, we obtain our final one-dimensional time-dependent Ginzburg-Landau theory,
\begin{equation}\label{}
\partial_t \varphi = \partial^2_z \varphi + \epsilon \varphi - g |\varphi|^2 \varphi + \xi .
\end{equation}


\begin{thebibliography}{99}
\bibitem{z_dependence}
Note that due to the translational invariance along the $z$-axis, the wavefunction should also contain a factor of $e^{ikz}$, which contributes $k^2$ to the energy eigenvalues. However, the ground state is always obtained for $k=0$, so we suppress the $z$-dependence in the wavefunctions.
\bibitem{gp_dynamics}
While we concentrate on relaxational dynamics, the general method outlined in Appendix~\ref{sec:dynamics} can be applied to other dynamical models, e.g., the reversible Gross-Pitaevskii dynamics of a Bose condensate.
\end{thebibliography}
\end{document}